# Democratizing Atomistic Simulation Workflows for the AI Era with the Quantum Accelerator

Andrew S. Rosen[1,*], Naisargi Goyal[1], Brad Ayers[2], Vineet Bansal[3], Julia H. Baratta[1], Samuel M. Blau[4], Yuan Chiang[5,6], Sihoon Choi[1], Orion Archer Cohen[7], Blake Dallmann[1], Tom Demeyere[8], Will Engler[9], Yue-Wen Fang[10], Isabella Furrick[1], Eliu Huerta[11,12], Honghui Kim[13], Hironori Kondo[14,15,a], Anup Kumar[16], Jaehong Kwon[1], Osman Mamun[17], Charles B. Musgrave III[18], Hananeh Oliaei[1], Aryan Saha[19], Davide Sarpa[2], Benjamin X. Shi[20], Yuliang Shi[1], Xing Wang[21], Robert B. Wexler[22]

[1]Department of Chemical and Biological Engineering, Princeton University, Princeton, New Jersey 08544, United States
[2]School of Chemistry and Chemical Engineering, University of Southampton, Southampton, United Kingdom
[3]Research Computing, Princeton University, Princeton, New Jersey 08544, United States
[4]Energy Technologies Area, Lawrence Berkeley National Laboratory, Berkeley, California 94720, United States
[5]Department of Materials Science and Engineering, University of California, Berkeley, California 94720, United States
[6]Materials Sciences Division, Lawrence Berkeley National Laboratory, Berkeley, California 94720, United States
[7]Riven Systems, New York, New York 10027, United States
[8]Department of Materials Chemistry, Federal Institute of Materials Research and Testing, 12205 Berlin, Germany
[9]Globus, University of Chicago, Chicago, Illinois 60637, United States
[10]Instituto de Nanociencia y Materiales de Aragón, CSIC-Universidad de Zaragoza, Zaragoza 50009, Spain
[11]Data Science and Learning Division, Argonne National Laboratory, Lemont, Illinois 60439, USA
[12]Department of Computer Science, The University of Chicago, Chicago, Illinois 60637, USA
[13]Graduate School of AI, Korea Advanced Institute of Science and Technology, Seoul, Republic of Korea
[14]Harvard John A. Paulson School of Engineering and Applied Sciences, Harvard University, Cambridge, Massachusetts 02138, United States
[15]Department of Chemistry and Chemical Biology, Harvard University, Cambridge, Massachusetts 02138, United States
[16]Quantum Software Engineering Team, D-Wave Quantum Inc., New Haven, Connecticut 06511, United States
[17]R. F. Smith School of Chemical and Biomolecular Engineering, Cornell University, Ithaca, New York 14853, United States
[18]Department of Materials Science and Engineering, Massachusetts Institute of Technology, Cambridge, Massachusetts 02139, United States
[19]Department of Electrical and Computer Engineering, Princeton University, Princeton, New Jersey 08544, United States
[20]Initiative for Computational Catalysis, Flatiron Institute, New York, New York 10010, United States
[21]Elemynt Pte. Ltd., 3 Fusionopolis Way, Singapore 138633, Singapore
[22]Department of Chemistry and Institute of Materials Science and Engineering, Washington University in St. Louis, St. Louis, Missouri 63130, United States
[a]Present address: Department of Materials Science and Engineering and Center for Computational Science and Engineering, Massachusetts Institute of Technology, Cambridge, Massachusetts 02139, United States
[*]Corresponding author: asrosen@princeton.edu

**Abstract**. We present the Quantum Accelerator (QuAcc), an open-source workflow library for atomistic simulations with an emphasis on quantum-mechanical calculations. QuAcc provides predefined workflow recipes spanning first-principles electronic-structure methods, semiempirical and tight-binding approaches, classical potentials, and foundation machine-learned interatomic potentials (MLIPs). A central design feature of QuAcc is its separation of domain-specific scientific logic from the workflow engine used to orchestrate and execute calculations. Workflows are written as ordinary Python functions and can be executed with multiple supported workflow engines without modifying the underlying source code, lowering the barrier to developing and contributing new workflows. QuAcc also streamlines the evaluation of foundation MLIPs by providing a unified platform for generating *ab initio* reference calculations consistent with the model of interest, mitigating methodological drift when assessing model performance. Together, these features make QuAcc a flexible and accessible framework for atomistic simulation workflows that have become central to the current era of machine learning and artificial intelligence.

## Introduction

Computational chemistry and materials science are in the midst of a data revolution, driven by significant advances in artificial intelligence (AI) as well as ongoing efforts to accelerate the materials design and discovery process.[1] Quantum-mechanical methods, such as those based on density functional theory (DFT), have played a central role in these efforts because they enable the prediction of physicochemical properties across a wide range of chemical systems and application areas. Countless machine learning models have been developed that rely on features from *ab initio* calculations as inputs for material property prediction or that seek to predict the results of *ab initio* calculations themselves.[2] In parallel, foundation machine learning interatomic potentials (MLIPs) have emerged as a new paradigm for computational chemistry and materials science, as they have made it possible to rapidly navigate potential energy surfaces at time and length scales that are intractable with *ab initio* methods alone.[3,4] While there are many challenges that must still be overcome to achieve the lofty goal of accelerating materials design and discovery,[5,6] quantum-mechanical simulation methods have an essential role in the ongoing AI era.

Given the increasingly widespread reliance on quantum-mechanical methods, there is a central need for open-source software that can provide access to high-quality, reproducible simulation workflows that build upon scalable workflow orchestration tools.[7] The computational chemistry and materials science communities are also becoming increasingly interdisciplinary, and quantum-mechanical modeling is no longer performed exclusively by specialists in electronic-structure theory. As a result, atomistic simulation software must make established methods accessible while providing sensible defaults and safeguards against erroneous calculations.[8] These requirements are becoming more important than ever as agentic AI systems become increasingly widespread for constructing and executing complex scientific workflows.[9–15] Even for the human user, it is impractical to be an expert in every quantum-mechanical simulation code and materials modeling task, which underscores the value of readily extensible and community-developed software libraries for carrying out complex atomistic simulation workflows.

For the purposes of this work, we will distinguish between three separate layers of computational infrastructure in atomistic simulations: workflow engines, atomistic simulation drivers, and workflow libraries. These categories are not always mutually exclusive, and some software packages provide functionality spanning multiple layers. A workflow engine is generally domain-agnostic and provides abstractions for defining, dispatching, and monitoring complex computational workflows. More than 300 workflow engines have been developed (though far fewer are actively maintained),[16] and it is unlikely that there will ever be a one-size-fits-all solution due to the diverse and conflicting needs of potential end users.[17] An atomistic simulation driver provides a unified interface to execute various kinds of simulation packages and to carry out common types of atomistic simulation tasks. Common examples include the Atomic Simulation Environment (ASE),[18] Pymatgen coupled to Custodian,[19] and TorchSim.[20] Finally, a workflow library is built around both a workflow engine and atomistic simulation driver to make it easy for users to execute, develop, and share computational chemistry and materials science workflows of varying degrees of complexity. Over the last several years, multiple software packages for computational materials science workflows have been developed, such as Atomate2[21,22] as well as those based around AiiDA[23,24] and Pyiron.[25] Most workflow libraries are closely coupled to a single workflow engine, which can hinder adoption if the user's computing needs do not align with the features offered by the supported workflow engine.

In this work, we present the main features and design philosophy behind the Quantum Accelerator (Figure 1), which we will refer to as QuAcc (pronounced "quack"). QuAcc is a workflow library for atomistic simulations that has been openly developed over the last five years. QuAcc has been applied to diverse areas of chemistry and materials science, including but not limited to surface science, molecular reactivity, the thermodynamics of solid-state materials, and superconductivity.[26–40] One of the central design objectives of QuAcc is to decouple the scientific logic from the workflow engine used to execute it. Atomistic simulation workflows in QuAcc are written as ordinary Python functions and can be executed with multiple supported workflow engines without the need to rewrite the underlying source code. As a result, the barrier for contributing or developing new workflows in QuAcc is greatly reduced since doing so does not require learning a given workflow language's custom syntax and its many nuances. QuAcc is built primarily around the community-adopted ASE library, such that all ASE-compatible energy/force calculators can be used. This combination of workflow engine interoperability and broad support for atomistic simulation codes is intended to make QuAcc easy to use regardless of the end user's computing needs.

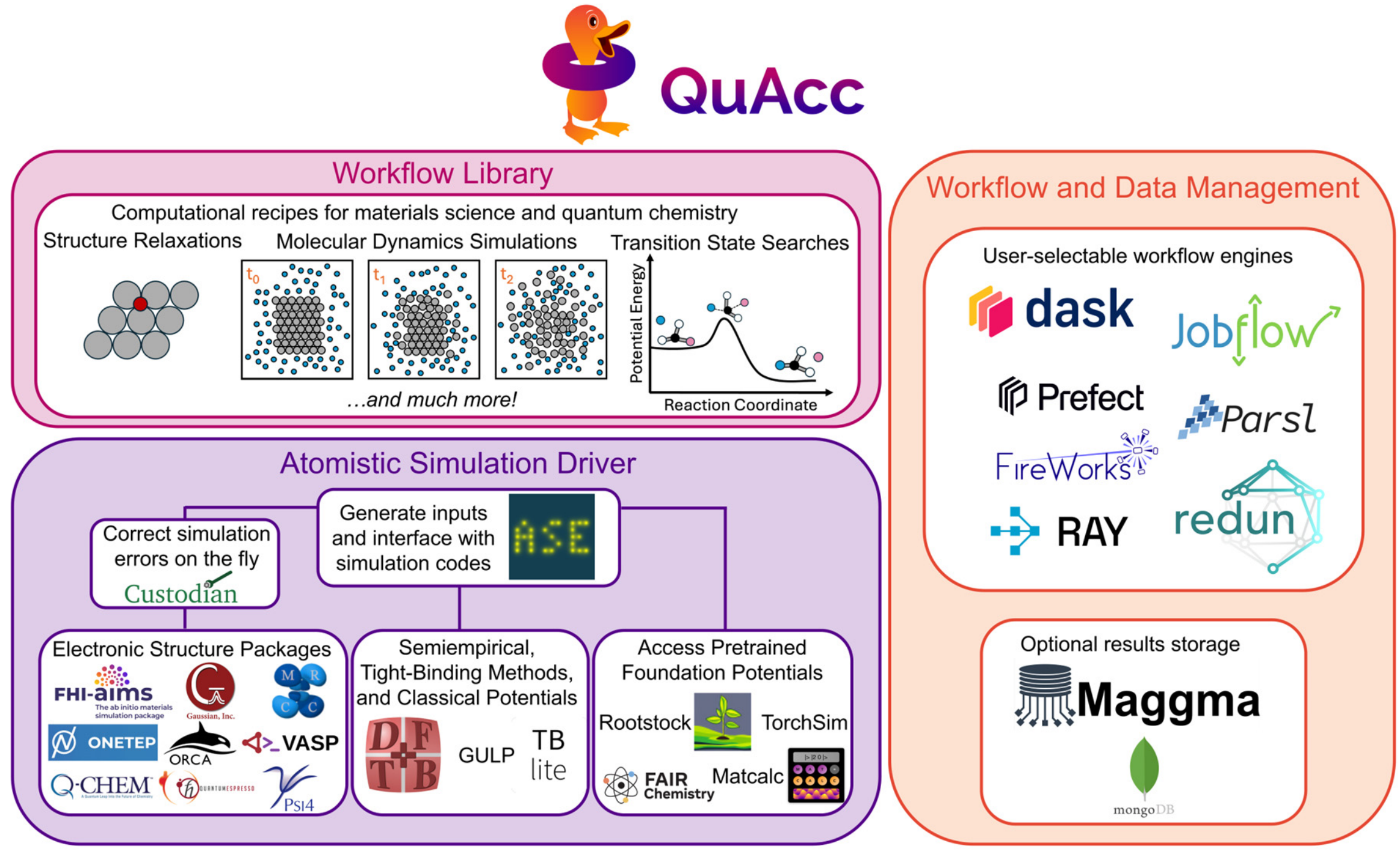

**Figure 1**. Overview of the Quantum Accelerator (QuAcc) and its components. The simulation codes for which there are existing recipes in QuAcc are highlighted, alongside the currently supported workflow engines.

### Library of Simulation Recipes

QuAcc provides a diverse library of predefined simulation "recipes" spanning first-principles electronic-structure methods, semiempirical methods, classical interatomic potentials, and MLIPs. At the time of writing, there are supported recipes for several electronic-structure packages, including FHI-aims,[41] Gaussian,[42] MRCC,[43] ONETEP,[44] ORCA,[45] Psi4,[46] Q-Chem,[47] Quantum ESPRESSO,[48,49] and VASP.[50,51] These interfaces collectively provide access to molecular and periodic DFT (including linear-scaling electronic-structure calculations) and correlated wave-function methods. The VASP and Q-Chem recipes benefit from additional functionality enabled by the Custodian library, which provides automated error handling for dozens of common errors across both codes. Additional recipes are available for semiempirical and tight-binding methods via DFTB+[52] and TBLite (which includes the xTB family of methods[53]) as well as classical potentials implemented via GULP,[54] including periodic GFN-FF.[55]

To ensure that QuAcc remains compatible with the rapidly expanding ecosystem of foundation MLIPs, the code can optionally interface with several external libraries—including Rootstock,[56] matcalc,[57] and FAIR-Chem[58]—that provide standardized access to a broad and continually evolving collection of pre-trained models accessible through ASE. Rootstock is a particularly useful integration component in QuAcc, as it lets the user define complex workflows using any supported MLIP without needing to manage the conflicting Python environments that many MLIPs require. The matcalc and FAIR-Chem libraries provide additional means of accessing pre-trained foundation MLIPs, although it is left to the user to install and manage the relevant dependencies. QuAcc also has a separate suite of TorchSim[20] recipes for calculations that benefit from batching multiple MLIP-based simulations on GPUs.

At a minimum, all supported simulation codes have recipes for static calculations and structure relaxations; however, most codes have additional recipes, such as those for molecular dynamics, surface calculations, vibrational frequency and phonon calculations, transition state searches, and band structure calculations. Adding new recipes for a given simulation code simply requires the developer to define a set of sensible default parameters for that job type and to stitch together existing recipes where appropriate. The recipe architecture is designed to remain extensible as new simulation methods and MLIPs become available. For ASE-based calculators beyond those shown in Figure 1, new recipes can be readily added by defining a suitable set of default arguments. In all cases, ASE is the main tool that is used to write any relevant input files and to interface with the underlying executable, where applicable. In select cases (e.g. Q-Chem, VASP), QuAcc defaults to having ASE make calls to Custodian,[19] which is a software package that identifies and corrects common simulation errors on-the-fly. In these cases, Custodian manages the job execution, but the input file writing and output parsing are still managed by ASE to provide a consistent user experience.

The output of each recipe in QuAcc is a dictionary that contains detailed information about the job metadata and important simulation results. We refer to these dictionaries as schemas. As with the workflow orchestration logic and the ASE-based calculator interface, QuAcc does not provide its own parsers for the underlying scientific simulation packages. By default, QuAcc uses ASE to write all input files and read key properties (e.g. structures, energies, forces, stresses) from the relevant simulation output files. In cases where mature file parsers exist that provide a richer set of (meta)data, these are also adopted in QuAcc as appropriate. For example, all VASP-based recipes in QuAcc build upon the schemas that are used as part of the Materials Project.[59]

**Architecture and Workflow Engine-Agnostic Design**

The workflow-engine-agnostic architecture of QuAcc allows users to select an orchestration system according to their unique computing needs without having to maintain different versions of the underlying QuAcc source code. At the time of writing, QuAcc supports the use of Dask (including integration with dask-jobqueue),[60] Jobflow (including execution through Jobflow-Remote and FireWorks[61]),[62] Parsl,[63] Prefect,[64] Ray,[65] and Redun[66] workflow engines as well as the option to forgo a workflow engine altogether to directly run the recipes as normal Python functions. As a result, QuAcc workflows can be run in a variety of environments ranging from a local workstation to high-performance computing clusters and distributed cloud-based resources. The workflow-engine-agnostic architecture also ensures that QuAcc can continue to function even if a particular workflow engine becomes unmaintained, as is common for open-source software.

While a thorough description of each supported workflow engine is beyond the scope of this work, we note that each option has its own unique features that may dictate whether an end user chooses to adopt it. Dask, Parsl, and Ray are designed for highly efficient task execution, making it possible to dispatch up to thousands of tasks per second across hundreds or even thousands of compute nodes; nonetheless, these codes lack the detailed job monitoring features of other workflow engines. Prefect is widely adopted in the data engineering community and offers substantial real-time monitoring capabilities, including visual dashboards to monitor large campaigns of jobs, but it has limited built-in support for monitoring complex workflows on supercomputers. Jobflow was developed by the Materials Project team and is designed for high-throughput simulations on academic supercomputers, relying heavily on the use of a database such as MongoDB. Finally, Redun is primarily meant for dispatching calculations on cloud computing resources, namely those based on Amazon Web Services (AWS).

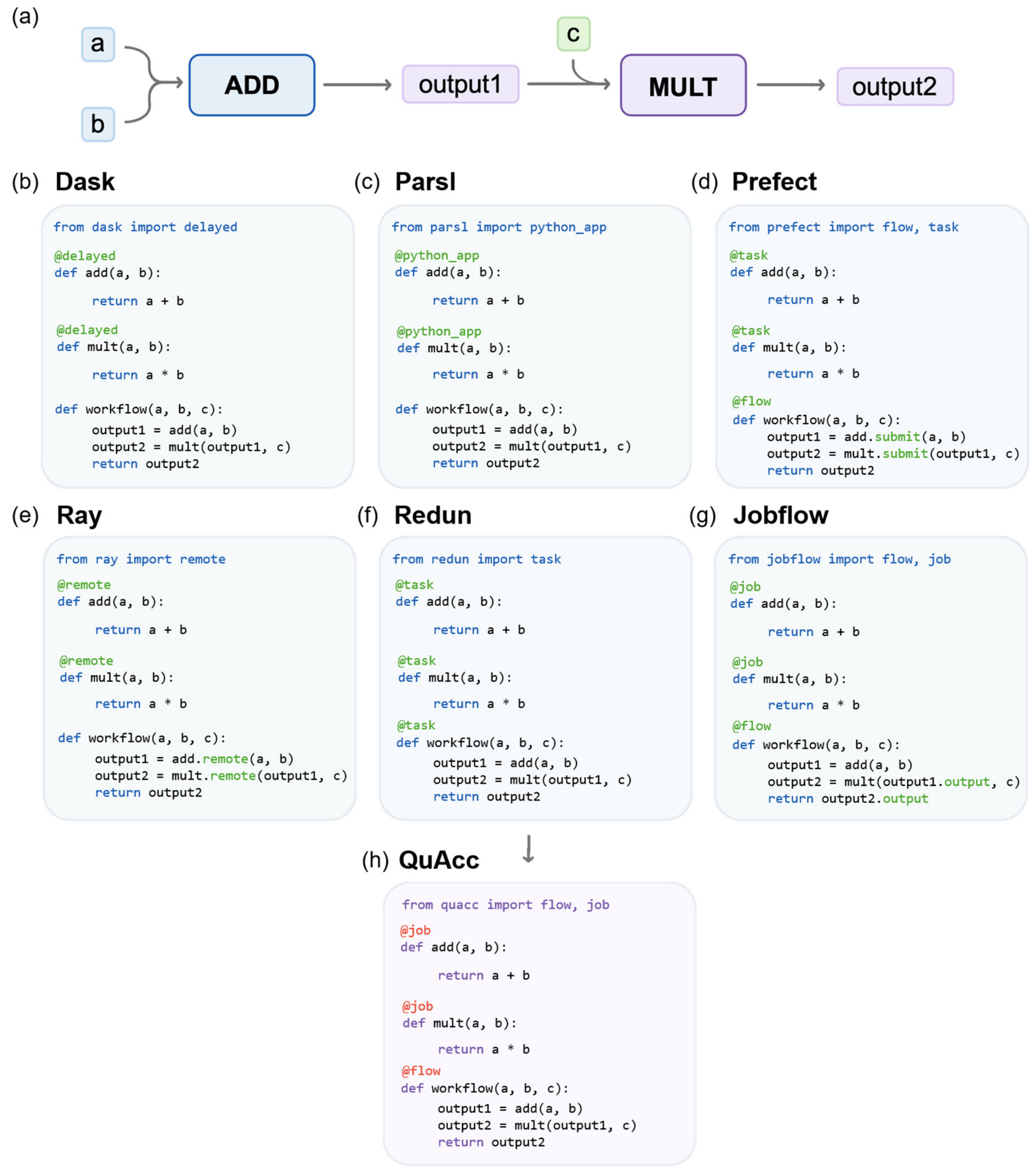

**Figure 2**. (a) A directed graph of a two-step workflow consisting of an addition job followed by a multiplication job. (b-g) The decorator-based syntax for different workflow engines that QuAcc supports. (h) The corresponding QuAcc decorators, which dynamically transform to the syntax of the corresponding workflow engine at import time based on the user's settings.

To achieve a high degree of interoperability between workflow engines, we built upon a convergent design philosophy that has emerged across many modern Python workflow engines: namely, the decorator-based programming model. With decorators, ordinary Python functions are transformed into independently executable tasks, and collections of tasks are assembled into multi-step workflows (Figure 2). Although the underlying execution details, data structures, and scheduling mechanisms differ among the various workflow engines, QuAcc exploits their similar syntax to provide a unified, workflow-engine-agnostic interface. In QuAcc, discrete computational tasks are defined using the @job decorator, whereas workflows that compose one or more jobs are defined using the @flow decorator. QuAcc also provides a @subflow decorator to denote more complex, dynamic workflow patterns where a collection of job outputs is returned, while the number of constituent jobs may not be known *a priori*. Apart from these decorators, the underlying Python functions are written in the same manner as conventional functions. As a result, the scientific logic of a recipe remains largely independent of the orchestration framework used to execute it.

When a decorated function is imported, QuAcc examines the user's settings to determine the selected workflow engine. QuAcc then dynamically maps its generic decorators onto the corresponding abstractions provided by that workflow engine. When no workflow engine is selected, the decorators are simply ignored, and the functions retain their ordinary, synchronous execution behavior. This design philosophy allows the same recipe to be developed and tested locally before being deployed through a distributed workflow system, without requiring any changes to the source code. Importantly, QuAcc delegates all aspects of job dependency management, task submission, monitoring, and resource allocation to the selected workflow engine and has no workflow logic of its own. This separation of responsibilities keeps the scientific recipes decoupled from infrastructure-specific orchestration logic and reduces the complexity of the codebase.

Despite the similar syntax of the supported workflow engines, several challenges had to be addressed to achieve the interoperability suggested by Figure 2. Consider a two-step quantum-chemistry workflow consisting of a structure relaxation followed by a static calculation where each job returns a dictionary of simulation results, and the second job accesses a specific key from the first job's output. This pattern is difficult for many workflow engines because the dictionary keys are not resolved until execution completes, and waiting for them would undermine the concurrent programming paradigm. Additionally, workflow engines differ in how they serialize and deserialize function arguments when transferring tasks between local and remote machines, with not all workflow engines supporting Python classes and domain-specific abstractions such as ASE Atoms objects. Addressing these limitations required both changes to the upstream repositories for the various workflow engines and to the QuAcc codebase.

**Dependencies and Testing**

To ensure that QuAcc is easy to install and use despite the large number of supported simulation methods, the default Python package ships with a minimal set of dependencies. Several optional dependency groups are defined depending on which recipes a user may be interested in running. Unit and functional tests are provided for the entire codebase, and QuAcc has near-100% line coverage at the time of writing. All tests can be run locally or automatically via GitHub Actions, the latter of which ensures that the core test suite is run for each pull request. For some simulation packages that can be readily installed via pip or the Conda package manager—such as DFTB+, Psi4, and Quantum ESPRESSO—real but inexpensive electronic structure calculations are run in the test suite. However, for simulation packages that cannot be installed as Python packages or are proprietary, the execution is mocked by default. For approved users and maintainers, pull requests on GitHub trigger a dedicated test suite to run on a local compute cluster that runs un-mocked tests (e.g. for Gaussian, Q-Chem, VASP) to ensure that the end-to-end pipeline remains functional.

**Representative Examples**

The recipes in QuAcc can be imported and used as-is or assembled into a custom workflow that the user defines. As a minimal toy example, we start by considering a two-step workflow consisting of a structure relaxation of a 2×2×2 supercell of bulk copper followed by a static calculation, which we carry out using the GFN2-xTB method[67] with periodic boundary conditions as implemented in the code DFTB+[52] (Figure 3a). Here, the two premade jobs were imported, and we have chained them together to make a custom workflow. The output is a dictionary containing the input parameters, updated ASE Atoms object, and the primary results of the simulation. By default, the workflow will run as standard Python code; however, by specifying one of the supported workflow engines via the QuAcc settings manager, the same code can be used to dispatch the workflow on a remote machine.

The recipes from different types of simulation methods can be easily combined in QuAcc as well. For example, Figure 3b demonstrates how one can carry out a relaxation with the TensorNet MLIP[68] trained on the MatPES-PBE dataset[69] followed by a static calculation in VASP using MatPES-compatible settings, as might be done to evaluate the energy, force, and stress errors of the MLIP-relaxed structure.

In addition to premade jobs, QuAcc ships with several premade workflows of varying degrees of complexity. As a representative example, Figure 3c demonstrates a dynamic workflow that takes a bulk structure as input, carves high-symmetry surface slabs, carries out a structure relaxation for each one, followed by a static calculation and the collection of all results. Here, we have chosen to use effective medium theory (EMT), simply for the ease of reproducibility by the reader. Since the relaxations of the surface slabs do not depend on one another, if a workflow engine is used, these jobs will be carried out concurrently, assuming there are available computing resources to do so.

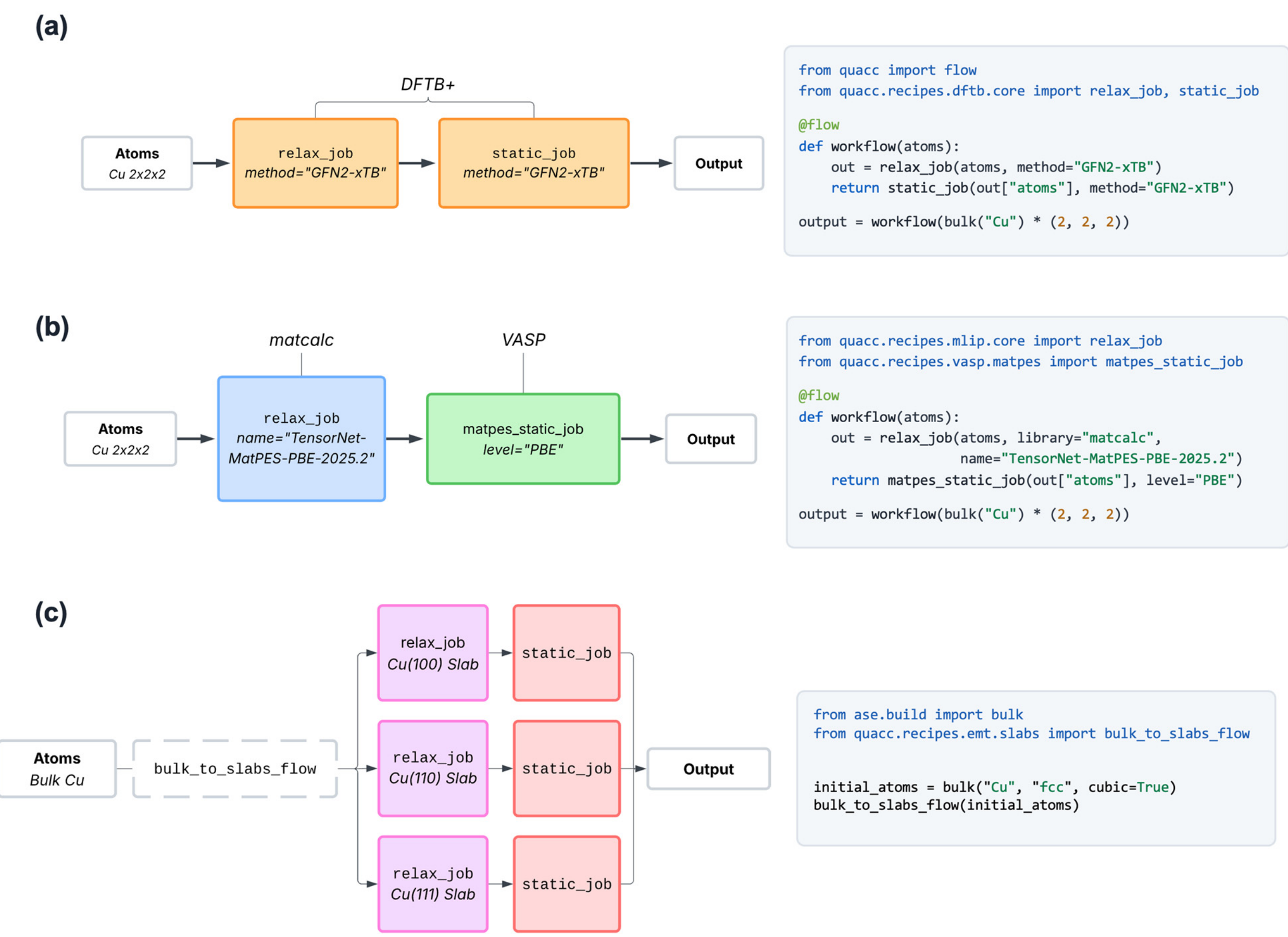

**Figure 3.** (a) Directed graph of a custom two-step workflow consisting of a GFN2-xTB relaxation followed by a corresponding static calculation. (b) Directed graph of a custom two-step workflow consisting of a structure relaxation using the TensorNet-MatPES-PBE machine learning interatomic potential followed by a static density functional theory calculation carried out with MatPES-PBE compatible settings. (c) Directed graph of a premade, dynamic workflow that generates unique surface slabs, carries out a relaxation on each using effective medium theory (for demonstration purposes) followed by corresponding static calculations.

## MLIP Training and Benchmarking

One of the many uses of QuAcc is for developing and evaluating large-scale foundation MLIPs. When evaluating the accuracy of pre-trained MLIPs for new materials or application areas, the accuracy of the energy (E), force (F), and (if applicable) stress (S) predictions is of central importance. While low E/F/S errors do not guarantee a model is accurate for downstream simulation tasks of interest, they are a critical prerequisite. In order to evaluate the E/F/S errors of a pre-trained MLIP, the general approach is to carry out single-point (i.e. static) calculations for representative configurations of interest and to compare the E/F/S predictions between the MLIP and the *ab initio* reference. Crucially, the *ab initio* calculations must be carried out using the same computational methodology as the data used to train the MLIP. Ideally, these calculations should use the same electronic-structure code, exchange–correlation functional, dispersion or Hubbard $U$ correction (if applicable), and choice of pseudopotentials or basis sets used in the original dataset. Differences in any of these settings can introduce major inconsistencies that might be incorrectly attributed to errors in the MLIP.

Reproducing a published computational protocol can be difficult for both specialists and nonspecialists alike, particularly when important numerical details are distributed across publications, input files, software repositories, and supplementary information. To streamline this process, QuAcc provides predefined DFT recipes designed to reproduce the computational settings used for several widely adopted MLIP datasets, including the Open Catalyst 2020, 2022, and 2025 (OC20, OC22, OC25),[70–72] Open Materials 2024 (OMat24),[73] Open Molecular Crystals 2025 (OMC25),[74] Open Molecules 2025 (OMol25) and Open Polymers 2026 (OPoly26),[75,76] Open Direct Air Capture 2023 and 2025 (ODAC23, ODAC25),[77,78] Materials Project Trajectory (MPtrj),[59,79,80] subsampled Alexandria (sAlex),[73,81] Materials Potential Energy Surface (MatPES),[69] and Materials Project – Active Learning of Off Equilibrium structures (MP-ALOE) datasets.[82] The same recipes can be used both to construct internally consistent benchmark sets and to generate additional training configurations that extend an existing dataset without introducing potential methodological incompatibilities. Due to the high-throughput infrastructure that QuAcc readily supports, *ab initio* training data can also be constructed with any ASE-supported calculator and choice of settings that the end user prefers. Finally, we note that QuAcc has several recipes to reproduce the settings of common *ab initio* datasets beyond those solely used for MLIP training, such as the Materials Project[59,79] and the Quantum Metal–Organic Framework (QMOF) Database.[83,84]

For the sake of demonstration, we ran single-point DFT calculations using QuAcc on 100 structures from the WBM high-energy states dataset,[85] which was obtained by running molecular dynamics (MD) simulations at 1000 K on structures from the WBM dataset.[86] The 100 structures cover 82 elements across the periodic table and were chosen to maximize elemental diversity, enabling us to examine how differences in pseudopotentials and in the treatment of Hubbard $U$ corrections across various elements may contribute to E/F/S errors. Five MLIPs were evaluated: TensorNet-MatPES-PBE (2025.2)[68,69] and MACE-MatPES-PBE-0,[87] which are trained or fine-tuned on the MatPES-PBE dataset;[69] and MACE-OMat-0 (medium), PET-OMat-XL (v1.0.0),[88] and UMA-OMat-s1p2p1,[58] which are trained or fine-tuned on the OMat24 dataset.[73] The MLIP predictions for energy, force, and stress were compared to single-point DFT calculations provided with the WBM high-energy states dataset (i.e. MPRelaxSet settings) and the corresponding MLIP training dataset settings as implemented in the pre-made QuAcc recipes.

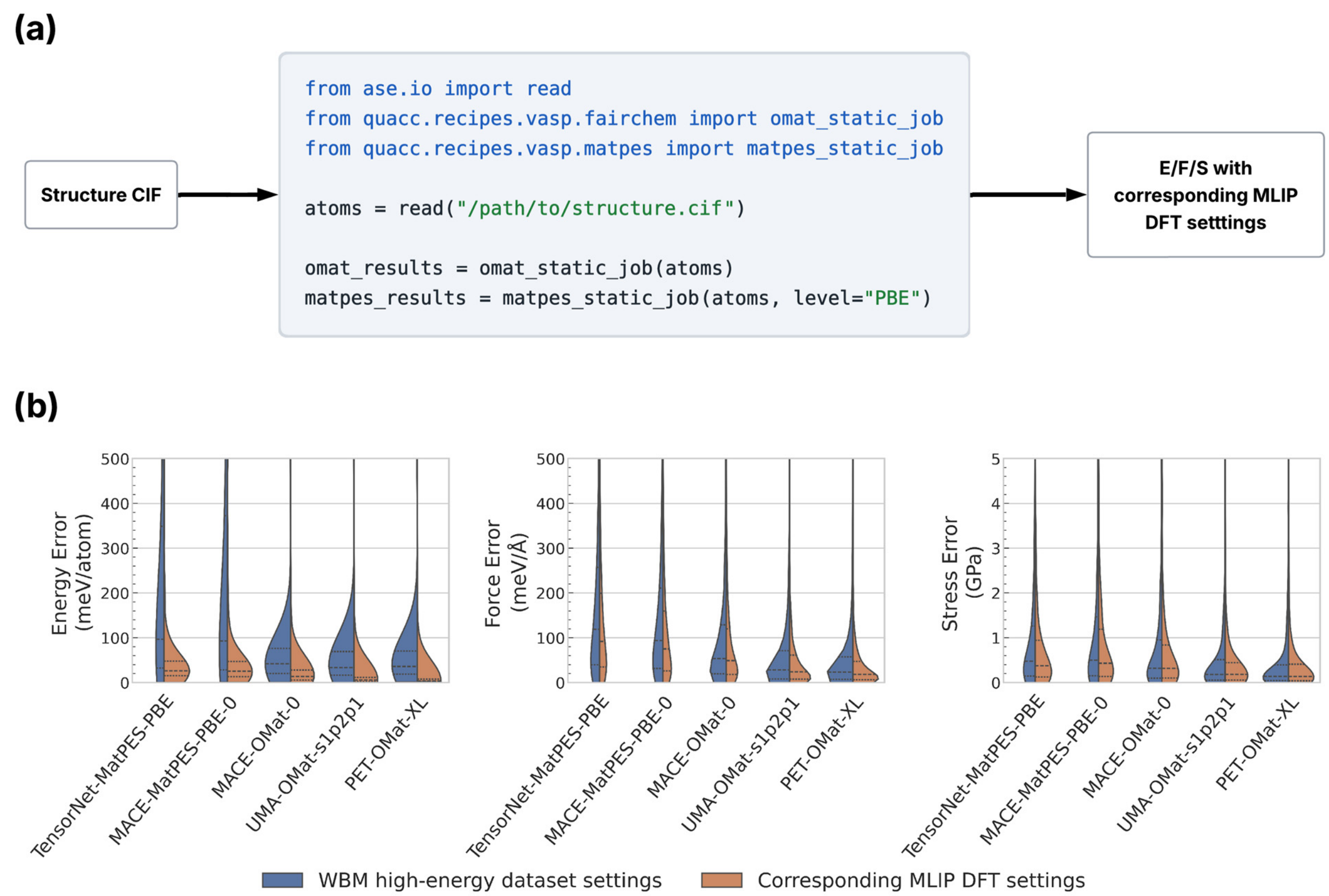

**Figure 4.** (a) Workflow for generating MLIP-consistent DFT ground truth using predefined QuAcc recipes. (b) Violin plots showing the energy, force, and stress errors for 100 structures from the WBM high-energy states dataset. The central dashed line represents the median, and the dotted lines represent the first and third quartiles. The *y*-axes are truncated for visual clarity.

**Table 1**. Energy, force, and stress mean absolute error (MAE) of five MLIPs evaluated against both the as-provided WBM high-energy states DFT labels and compatible DFT settings for the given MLIP obtained with QuAcc.

| Model | DFT Settings | Energy MAE (meV/atom) | Force MAE (meV/Å) | Stress MAE (GPa) |
|---|---|---|---|---|
| TensorNet-MatPES-PBE | WBM | 260.8 | 191.5 | 0.984 |
| | MatPES-PBE | 50.4 | 146.0 | 0.783 |
| MACE-MatPES-PBE-0 | WBM | 261.7 | 168.1 | 1.134 |
| | MatPES-PBE | 49.7 | 117.8 | 0.879 |
| MACE-OMat-0 | WBM | 70.1 | 105.2 | 0.756 |
| | OMat24 | 31.2 | 90.1 | 0.739 |
| UMA-OMat-s1p2p1 | WBM | 69.2 | 63.9 | 0.489 |
| | OMat24 | 22.2 | 47.2 | 0.497 |
| PET-OMat-XL | WBM | 69.4 | 55.3 | 0.405 |
| | OMat24 | 19.8 | 38.0 | 0.432 |

As seen in both Figure 4 and Table 1, the choice of DFT reference settings has a significant effect on apparent MLIP accuracy. MatPES-trained models exhibit large apparent energy errors relative to the as-provided WBM high-energy states DFT labels. In contrast, the same MLIP predictions evaluated against fully compatible DFT settings (i.e. those used to generate the MatPES dataset) yield a fivefold reduction in the energy MAE. Similar trends are observed with MLIPs trained on the OMat24 dataset, where energy MAE decreases by more than a factor of two when compared against ground-truth labels generated using OMat24 DFT settings with QuAcc. The particularly large energy discrepancy observed for MatPES-trained MLIPs is due in part to the absence of Hubbard $U$ corrections in the MatPES dataset, whereas both the OMat24 and WBM datasets include Hubbard $U$ corrections for select material compositions.[89] Force MAEs are also impacted by the choice of DFT settings, with the apparent MAE decreasing by roughly 15–30% across all five models when compared against the fully consistent DFT settings instead of the WBM high-energy states DFT labels. Stress errors, in contrast, show smaller and less consistent changes across models. We note that both the MPRelaxSet-based DFT labels in the WBM high-energy states dataset and those of the OMat24 dataset rely on a self-consistent field energy convergence criterion of $5.0\times10^{-5}$ eV/atom, which can affect numerical convergence, particularly for the unit cell stresses of larger structures.

The results of this exercise demonstrate that inconsistencies in the DFT settings used to evaluate MLIPs can artificially inflate MLIP errors. Therefore, meaningful MLIP benchmarking requires that the ground-truth labels for single-point DFT calculations be generated using the same methodology as the training data for the MLIP (or the fine-tuning dataset for fine-tuned models). QuAcc's predefined recipes facilitate the benchmarking of several pre-trained MLIPs against internally consistent reference data, hence allowing the reported errors to more reliably reflect the MLIP performance rather than discrepancies arising from differences in the underlying DFT methodology.

**Outlook**

The scientific software landscape in chemistry and materials modeling is ever evolving, and it is worth reflecting on the state of the field and what the future may hold for workflow libraries such as QuAcc. The scientific community has now reached a point where high-throughput calculations are routine, enabled in large part by well-established workflow libraries such as QuAcc and related codes including Atomate2,[21] AiiDA,[22,23] and Pyiron.[25] At the same time, the field is approaching another inflection point in which large language models (LLMs) coupled with agentic AI methods are likely to play an increasingly important role in both scientific software development and its use. This transition creates new opportunities for automating complex chemistry and materials modeling studies, but it also places new demands on the underlying software infrastructure.

Widely adopted LLMs are trained on broad corpora that do not necessarily reflect high-quality, expert-designed computational protocols. In this context, workflow libraries like QuAcc can serve as an important layer between an AI agent and the underlying simulation software. Rather than requiring an agent to independently reconstruct every input setting from scratch, the agent can invoke established recipes that encode sensible defaults and best practices. In fact, the AI coding agent platform CatGo has already demonstrated this vision by importing and building upon QuAcc templates.[10] The workflow-engine-agnostic design of QuAcc may also become increasingly important as agentic systems are applied across heterogeneous computing environments. The same workflow can be constructed, inspected, and tested locally and then dispatched to a remote machine via a suitable workflow engine without rewriting the code itself.

We expect the role of workflow libraries like QuAcc to expand rather than diminish as scientific AI systems become more capable. LLM-based agents may increasingly determine what calculations should be performed, assemble workflows from composable recipes, analyze intermediate results, and adapt subsequent calculations in response to new findings. However, domain-specific workflow libraries will remain essential for defining how those calculations can be carried out reliably and providing natural guardrails. As a result, the continued development of open, interoperable, and community-maintained workflow libraries like QuAcc will remain important for ensuring that increased automation in computational chemistry and materials science is accompanied by equally strong standards for reproducibility, transparency, and scientific reliability.

## Conclusion

In this work, we present the Quantum Accelerator (QuAcc), an open-source workflow library for atomistic simulations designed to make computational chemistry and materials modeling workflows more accessible, reproducible, and interoperable with various workflow engines. QuAcc provides a broad collection of predefined simulation recipes spanning first-principles electronic-structure methods, semiempirical and tight-binding approaches, classical interatomic potentials, and foundation MLIPs. By building primarily upon ASE and supporting multiple workflow engines through a unified syntax, QuAcc separates the scientific logic of a simulation workflow from the infrastructure used to execute it. This design allows the same underlying recipes to be run locally, on high-performance computing resources, or in distributed environments without requiring users to rewrite the scientific workflow for a particular orchestration framework.

Beyond simplifying routine atomistic simulations, QuAcc provides infrastructure for constructing reproducible datasets based on high-throughput calculations and for benchmarking foundation MLIPs against consistently defined *ab initio* reference calculations. Its extensible recipe architecture, workflow-engine-agnostic design, and use of ordinary Python functions are intended to lower the barrier for both using and developing complex computational workflows. As atomistic simulations become increasingly intertwined with machine learning and agentic AI systems, we anticipate that interoperable workflow libraries such as QuAcc will provide an important foundation for translating scientific objectives into reliable and reproducible calculations. In this way, QuAcc aims to help democratize access to high-quality atomistic simulation workflows (for both humans and machines) while remaining flexible enough to evolve alongside the rapidly changing computational chemistry and materials science ecosystem.

## Data Availability

The QuAcc code can be found at https://github.com/Quantum-Accelerators/quacc. Each version is archived on Zenodo at the following DOI: https://doi.org/10.5281/zenodo.7720998.

## Author Contributions

A.S.R. designed and led the project, wrote the majority of the QuAcc code, maintains the codebase, and wrote the manuscript with input from the co-authors. N.G. carried out the MLIP and DFT calculations in this work and drafted the corresponding section. J.H.B., I.F., N.G., and A.S. made the figures in the manuscript. The remaining co-authors contributed code to the openly accessible software repository. The code contributors were not directly involved with the National Science Foundation-funded project(s) described in the Acknowledgments section unless explicitly stated.

## Acknowledgments

A.S.R., I.F., and N.G. acknowledge support from the National Science Foundation under Award No. OAC-2514141. E.H. and W.E. acknowledge support from the National Science Foundation under Award No. OAC-2514142. S.M.B. acknowledges support from the Center for High Precision Patterning Science (CHiPPS), an Energy Frontier Research Center funded by the U.S. Department of Energy (DOE), Office of Science, Basic Energy Sciences (BES). Y.-W.F. acknowledges support from the national research project PID2025-173011NA-I00 and from the Severo Ochoa Centre of Excellence programme CEX2023-001286-S, both funded by MICIU/AEI/10.13039/501100011033. The Flatiron Institute is a division of the Simons Foundation.